\documentclass[a4paper, 10pt]{article}

\usepackage{graphics}
\usepackage{graphicx}

\usepackage{amsmath,amsfonts,amssymb,amsthm}

\usepackage{newtxtext}
\usepackage[varvw]{newtxmath}

\def\R{{\mathbb{R}}}

\title{Action potential solitons and waves in axons}

\author{Gaspar Cano$^{1,2}$ and  Rui Dil\~ao$^{2,3}$}
\date{\today}

\begin{document}

\maketitle

\begin{center}
1-Department of Biology, Institute for Theoretical Biology, Humboldt-Universit\"at zu Berlin, Berlin, 10115, Germany, Berlin, 10115, Germany\\
2-University of Lisbon, Instituto Superior T\'ecnico, 
Av. Rovisco Pais, 1049-001 Lisbon, Portugal\\
3-Institut des Hautes \'Etudes Scientifiques, 
 35, route de Chartres, 91440 Bures-sur-Yvette, France
 \end{center} 
 
\begin{center}
gaspar.fcano@gmail.com\\
ruidilao@tecnico.ulisboa.pt
\end{center}

\begin{abstract}
We show that the action potential signals generated inside axons are reaction-diffusion solitons or reaction-diffusion waves, refuting the Hodgkin and Huxley hypothesis that action potentials propagate along axons with an elastic wave mechanism. 
Reaction-diffusion action potential wavefronts and solitons annihilate at collision and boundaries of axons, in contrast with elastic waves, where amplitudes add up and reflect at boundaries. 
We numerically calculate the values of the speed of the action potential spikes and the dispersion relations.  
These findings suggest several experiments as validating and falsifying tests for the Hodgkin and Huxley model.
\end{abstract}


\section{Introduction}\label{sec1}

The electrophysiological states of cells and axons are characterised by an electric potential drop across the cellular membrane, maintained through the exchange of ions between the cytoplasm and the intercellular space, \cite{Oh}, \cite{Pur}, and \cite{Phi}. To describe the electrical properties of axonal signalling, in a sequence of papers, Hodgkin and Huxley (HH) introduced a mathematical model aiming to describe the propagation of action potentials in the axoplasm. Which they have compared with voltage-clamp data taken from the axon of the squid \textit{Loligo}, \cite{HH}. 

In current-clamp experiments, one of the electrodes is located in the extracellular space and the second one is a thin wire introduced longitudinally into the axon, \cite{Tas}, \cite[p. 143]{Leu1}. When the axon is electrically excited away from the inner electrode, the measured electrical potential drop is a spiky (negative) signal that evolves in time, \cite[p. 24]{Hil}. In principle, this signal results from a longitudinally propagating signal --- the action potential --- measured by the inner electrode inside the axon.

The derivation of the HH mathematical model for the action potential phenomenon relies on the analogy between the potential difference measured on both sides of the cellular membrane and an electric circuit containing a variable resistance and a power source in series, both in parallel with a capacitor. This analogy is phenomenological, aiming to explain the gating mechanism of ion channels across a cellular membrane through a variable resistance. The power source and the capacitor describe a source of energy and a potential energy storage reservoir. The biological functions of the three electric components are unspecified, \cite[p. 152]{Tas} and \cite[p. 152]{Leu1}.

In a synthetic form, the HH model equations are
\begin{equation}
\left\{
\begin{array}{rcl} \displaystyle
{\tilde D}{\frac{\partial^2 V}{\partial x^2}} &=&\displaystyle C\frac{\partial V}{\partial t}+F(V,\vec n)+i(x,t)\\ [8pt]\displaystyle
\frac{\partial \vec n}{\partial t} &=&\displaystyle \vec G(V,\vec n),
\end{array}\right.
\label{eq1}
\end{equation}
where $t$ is time, $x$ is the position coordinate along the axon, $V(x,t)$ is the potential drop across the cellular membrane, ${\tilde D}$ is a diffusion coefficient of the potential drop along the axoplasm, $C$ is the phenomenological capacitance of the axon membrane, and $i$ is a current eventually describing an external forcing, as in current-clamp experiments, or simply a neuronal signal originated in the main body of a neurone. 
The vector function $\vec n (x,t)=(n,m,h)$ contains gating variables, specific to ion types.  
The functions $F(V,\vec n):\R^4\to \R$ and $\vec G(V,\vec n):\R^4\to \R^3$ describe, respectively, the local response to the potential drop changes across the cellular membrane and the gating mechanisms of ion channels, \cite{Rin}, \cite{Has}, \cite{Erm}, \cite{KS} and \cite{HP}.

 Hodgkin and Huxley conjecture that the propagation properties of the action potential are analogous to those of a propagating elastic wave, \cite[p. 522]{HH}. They assumed the existence of an unknown mechanism which would impose an elastic wave-type propagation mechanism inside the axon, such that the transmembrane potential would propagate according to the wave equation
\begin{equation}
{\frac{\partial^2 V}{\partial x^2}} =\frac{1}{\theta^2} {\frac{\partial^2 V}{\partial t^2}},
\label{eq2}
\end{equation}
where $\theta $ is an unknown {\it ad hoc}  speed constant. Even though the solutions of equations \eqref{eq1} and \eqref{eq2} are generically incompatible (equation \eqref{eq1} is of parabolic type, and equation \eqref{eq2} is of hyperbolic type),  Hodgkin and Huxley have substituted the first term in \eqref{eq1} by the second term in \eqref{eq2}, obtaining
\begin{equation}
\left\{
\begin{array}{rcl} \displaystyle
\frac{1}{\theta^2} {\frac{d^2 V}{d t^2}}&=&\displaystyle \frac{C}{{\tilde D}}\frac{d V}{d t}+\frac{1}{{\tilde D}}F(V,\vec n)+\frac{1}{{\tilde D}}i(x,t)\\ [8pt]\displaystyle
\frac{d \vec n}{d t} &=&\displaystyle \vec G(V,\vec n),
\end{array}\right.
\label{eq3}
\end{equation}
which is an ordinary differential equation. Hodgkin and Huxley integrated the ordinary differential equation \eqref{eq3} numerically for guessed choices of the free parameter $\theta$ and compared them with voltage-clamp data. Even though they obtained numerical results similar to their experimental data for some values of the chosen parameters, it is clear that an ordinary differential equation doesn't describe a time-dependent spatial phenomenon such as the propagation of an electric signal along the axon.

Besides the mathematical inconsistency just described, from the physical and biological points of view, there is a lack of the specific biochemical mechanisms that lead to the electric analogue of the model, as evidenced by experimental data, \cite{Leu1}. Several authors, based on physical and chemical principles, provided plausible evidence of the inadequacy of the HH model, \cite{Cl}, \cite{Meu}, \cite{Leu2}, and \cite{Bez}. For a recent review of the discussion on the validity of the HH model, we refer to \cite{Pey}.

These simple remarks show that model equations \eqref{eq1} can eventually describe action potential propagation. However, the argument for its calibration based on equations \eqref{eq3} should be reformulated.

The diffusion free HH model (${\tilde D}=0$ in \eqref{eq1}) has been extensively analysed from the point of view of its bifurcations,  \cite{Rin}, \cite{Has}, \cite{Erm} and \cite{CD}. These approaches have been used to obtain simplified models where parameters lose some biological meaning. These simplified models cannot produce predictions about the spatial propagation of action potentials, nor can the existence of Hopf bifurcations predict the generation of action potential signals. In \cite{CD}, it was shown that action potentials originated from a type I intermittency phenomenon (\cite{Pom}) associated with a saddle-node homoclinic bifurcation of limit cycles, which does not exist near Hopf bifurcations.

Some propagation properties of action potential signals have been studied in \cite{AM}. These authors have analysed localised high amplitude perturbations of the transmembrane potential of the HH equation and, by manipulating the initial distribution of the membrane potential along the axon, found that action potential fronts propagate as solitary waves, identifying the collisional annihilation of action potential spikes. This has been analysed for several values of the potassium Nernst potential $V_{\hbox{K}}^N$, with $i=0$. These findings are essential for understanding the fluctuation dynamics of the transmembrane potential but are challenging to observe in axons with voltage-clamp experimental techniques. 

More recently, due to the solitary characteristics of action potential signals observed by Hodgkin and Huxley, several authors attempted to explain the solitonic properties of the action potential as a non-linear elastic wave similar to the Korteweg-de Vries equation (Eq. (9.14) in \cite{App}). This approach is independent of the HH ionic mechanism and has no connection with biological parameters. On the other hand, solitary elastic waves of the Korteweg-de Vries type are characterised by different mechanisms of interaction and reflection at spatial boundaries, \cite{ZK}. As we will show below, the waves generated by the HH model equations are reaction-diffusion waves with different interaction laws. These other interaction characteristics also appear in chemical kinetics models with reaction-diffusion waves, \cite{SD} and \cite{DV}.

This paper characterises reaction-diffusion waves, reaction-diffusion solitary waves, and reaction-diffusion solitary wave packets in the HH model. The qualitative properties of the HH model and the comparison with experimental data test the model's validity and falsifiability. 

This paper is organised as follows. In section~\ref{sec2}, we review some of the results of the HH partial differential equation model \eqref{eq1} and some of the properties of its solutions, \cite{CD}. We also define the parameter settings of the model and summarise the numerical set of simulations. In section~\ref{sec3}, we show that the HH model equations \eqref{eq1} have reaction-diffusion type solitonic and oscillatory solutions, behaving as solitary waves or as solitary wave packets in the intermittent dynamical regime of the diffusion free equation \eqref{eq1}.  We derive the interaction properties of this type of reaction-diffusion wave and calculate wave speeds and dispersion relations of asymptotic regimes. Finally, in section~\ref{sec4}, we propose several experiments to validate the HH model qualitatively, and we summarise the paper's main conclusions.

\section{Action potentials propagate as reaction-diffusion waves}\label{sec2}
  
In \cite{CD}, we have exhaustively analysed the solutions of the HH partial differential equations \eqref{eq1}, in a spatial one-dimensional domain $I=[0,L]$, with $L<\infty$, and with zero flux boundary conditions. We have chosen the current function: $i(x,t)=i_0$, for $x=0$ and every $t\ge 0$, and $i(x,t)=0$, otherwise, and 
we have done the bifurcation analysis of the solutions of the reaction-diffusion equation \eqref{eq1}, as a function of the diffusion coefficient ${\tilde D}$ and of the parameter $i_0$. This particular choice of the external function $i(x,t)$ simulates current-clamp experiments. For the calibrated parameters,  propagating action potentials and action potential wave packets are generated near the left boundary of an axon. Some major conclusions derived from the HH model are important to recall:
\begin{itemize}
\item[1)] For positive but small values of $i_0$, the shape of the action potentials is caused by a type I intermittent response of the HH equations associated with a saddle-node bifurcation of limit cycles of the diffusion free system of equations (${\tilde D}=0$). This particular response is caused by a transient process that anticipates a transition from a dynamics with a unique stable steady-state to a dynamics with two limit cycles, one stable and the other unstable. These isolated spikes propagate without attenuation and therefore are solitary waves or solitons. 
Near the bifurcation, it is possible to obtain single action potential spikes or packets of propagating action potential spikes, depending on the intensity of the perturbation $i_0$.
\item[2)] For ${\tilde D}>0$ and larger values of $i_0$ when compared with case 1),  we found propagating periodic waves of action potentials spikes. We have numerically measured the propagation speeds depending on the model's parameters. This propagation speed is not an external parameter, as assumed in equations \eqref{eq2} and \eqref{eq3}.
\item[3)] For ${\tilde D}>0$ and larger values of $i_0$ when compared with case 2), we found solutions behaving chaotically and solutions with a long chaotic transient, which, as time passes, converge to a steady homogeneous state (chaotic or type III intermittency) --- dynamic summary in figure~\ref{fig:regions}.
\item[4)] Action potential propagation phenomenon only occurs if the current stimulus $i_0$, at the $x=0$ boundary of the axon, is large enough and persists long enough. This requirement is essential for forming the action potential at the current injection point. Once this happens, even if $i_0$ is set to zero during the spatial propagation, the action potential will reach the right boundary of the axon without attenuation.
\end{itemize}

All these properties of the HH partial differential equation model (\ref{eq1}) are predictions about the dynamics of action potentials and should be used to validate the HH model. 

To test the above-mentioned dynamic features of the HH model, we use the following parameterisation of equation \eqref{eq1}  
\begin{equation}
\begin{array}{rcl}\displaystyle
C\frac{\partial V}{\partial t}&=&\displaystyle {\tilde D}{\frac{\partial^2 V}{\partial x^2}}-g_{\hbox{Na}} m^3h(V-V^N_{\hbox{Na}})\\\displaystyle&&\displaystyle
-g_{\hbox{K}}n^4(V-V^N_{\hbox{K}})-g_{\hbox{L}}(V-V^N_{\hbox{L}}) -i\\ \displaystyle
\frac{\partial n}{\partial t} &=&\displaystyle \alpha_n(V)(1-n)-\beta_n(V)n =G_n(V,n)\\ [8pt]\displaystyle
\frac{\partial m}{\partial t} &=&\displaystyle \alpha_m(V)(1-m)-\beta_m(V)m =G_m(V,m)\\ [8pt]\displaystyle
\frac{\partial h}{\partial t} &=&\displaystyle \alpha_h(V)(1-h)-\beta_h(V)h =G_h(V,h),
\end{array}
\label{eq10}
\end{equation}
where
\begin{equation}
\begin{array}{lclrcl}\displaystyle
\alpha_n &=&\displaystyle 0.01 \phi \frac{V+10}{e^{(V+10)/10}-1}, &\displaystyle \beta_n &=&\displaystyle 0.125 \phi e^{V/80},\\ \displaystyle
\alpha_m &=&\displaystyle 0.1 \phi \frac{V+25}{e^{(V+25)/10}-1}, &\displaystyle \beta_m &=&\displaystyle 4 \phi e^{V/18},\\ \displaystyle
\alpha_h &=&\displaystyle 0.07 \phi e^{V/20}, &\displaystyle  \beta_h &=&\displaystyle \phi \frac{1}{e^{(V+30)/10}+1},\\ \displaystyle
\phi &=&\displaystyle 3^{(T-6.3)/10}. &&&
\end{array}
\label{eq11}
\end{equation}
In this equation, $V$ is the potential transmembrane drop, measured in mV, $i$ is a transmembrane current density injected into the axon, measured in $\mu$A/cm$^2$, and time is measured in ms.  
Positive values of $i$ correspond to currents from outside to inside the axon. 
 In equation (\ref{eq10}), the membrane potential is defined following the original HH paper, \cite{HH}, where the action potential voltage spikes are negative.
The gating variables $n$, $m$ and $h$ describe the opening and closing of the channel gates, are specific to the ion type and are dimensionless. %
The ionic conductances across the cellular membrane are
$g_{\hbox{Na}}$ and $g_{\hbox{K}}$, and $g_{\hbox{L}}$ is a constant measuring ``leak" conductance. $C$ is the membrane capacitance, and ${\tilde D}$ is a constant inversely proportional to the resistance ($\Omega$), measured along the axon of nerve cells. 

The model equations \eqref{eq10}-\eqref{eq11} have been calibrated for the squid giant axon at the  temperature $T=6.3$~$^\circ$C, \cite{HH}, and the values of the parameters are: $C=1$~$\mu$F/cm$^2$, $g_{\hbox{Na}}=120$~mS/cm$^2$, $g_{\hbox{K}}=36$~mS/cm$^2$ and $g_{\hbox{L}}=0.3$~mS/cm$^2$, where S=$\Omega^{-1}$ (siemens) is the unit of conductance. The Nernst equilibrium potentials, relating the difference in the concentrations of ions 
 between the inside and the outside of cells with the transmembrane potential drop, are $V^N_{\hbox{Na}}=-115$~mV, $V^N_{\hbox{K}}=12$~mV and $V^N_{\hbox{L}}=-10.613$~mV. 
 This choice of parameters is rescaled so that, at rest ($i=0$), the steady state of the transmembrane potential is $V^*(0)=0$~mV.
Hodgkin and Huxley have shown that the transmembrane diffusion coefficient is ${\tilde D} = a/(2R_2)$, where $a$ is the radius of the axon (considered as a cylinder) and $R_2$ is the specific resistivity along the interior of the axon. For the case of the squid giant axon, $a = 238$~$\mu$m, $R_2 = 35.4$~$\Omega~$cm and ${\tilde D}=3.4\times 10^{-4}$~S. 

To validate the Hodgkin-Huxley model predictions, we simulate the solutions of the reaction-diffusion equation (\ref{eq10}) in a domain of length $L=100$~cm, with zero flux boundary conditions. The spatial region has been divided into $M=800$ small length intervals $\Delta x$, where $L=M\Delta x$. 
We have used an explicit numerical method minimising the global error of the solution \cite{dilao}. Let $D$ be the diffusion coefficient, given by $D=\tilde{D}/C$. With the minimising condition 
$D=\Delta x^2/(6\Delta t)=L^2/(6 M^2\Delta t)$, and the choice 
of the diffusion coefficient $D=0.34$~cm$^2/$ms, or ${\tilde D}=3.4\times 10^{-4}$~S, in agreement with the value suggested by Hodgkin and Huxley \cite{HH}, we obtain the time step $\Delta t=0.00765931$~ms.

\section{Results}\label{sec3}

\subsection{Overall behaviour of the HH equations}

\begin{figure}[h]
\begin{center}
\includegraphics[width=0.95\hsize]{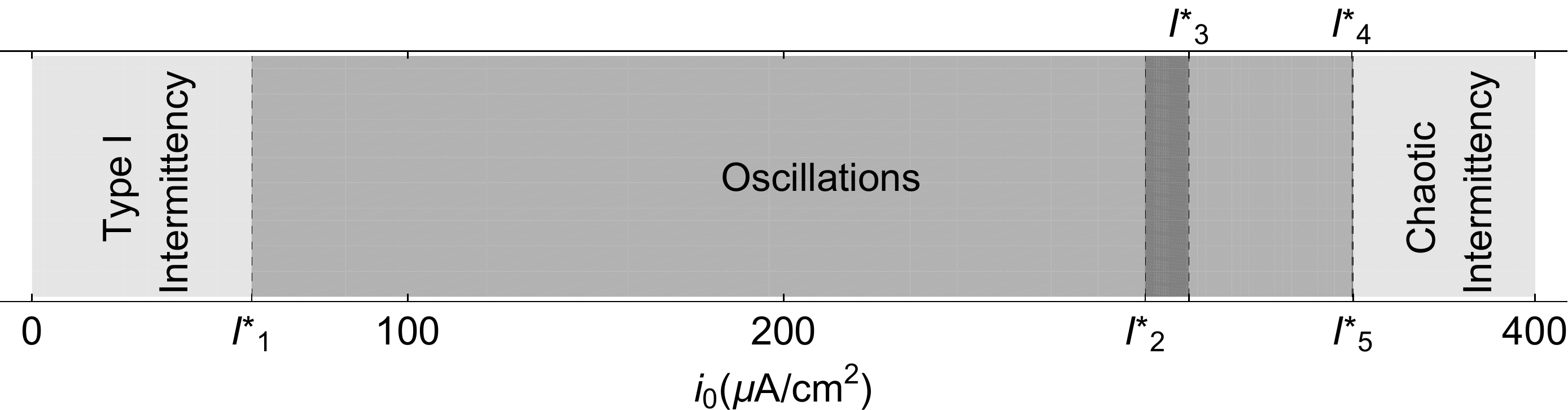}
\end{center}
 \caption{Bifurcation diagram of the solutions of the HH equations (\ref{eq10})-(\ref{eq11}), for ${\tilde D}=3.4\times 10^{-4}$~S, as a function of the parameter $i_0$. For this parameterisation, $I^*_1=58.44$, $I^*_2=296.23$, $I^*_3=307.85$, $I^*_4=351.21$ and $I^*_5=351.55$.} 
 \label{fig:regions}
\end{figure}

If an axon is initially at rest ($V=0$, for all $x$), it can be perturbed through the transmembrane current $i(x,t)=i_0$, for $x=0$ and every $t\ge 0$, and $i(x,t)=0$, otherwise. In figure~\ref{fig:regions}, we show 
a bifurcation diagram of the HH model as a function of $i_0$ for a chosen diffusion coefficient.

For $i_0 < I^*_1$, the system responds with type I intermittency, generating a finite number of action potential spikes. For $i_0 \in [I^*_1,I^*_5]$, the system oscillates indefinitely, never returning to rest. For $i_0 \in [I^*_1,I^*_2] \cup [I^*_3,I^*_4]$, asymptotically in time, the oscillations converge to a periodic solution. For the small regions $[I^*_2,I^*_3]$ and $[I^*_4,I^*_5]$, the oscillations show chaotic behaviour with period bifurcations. For $i_0 > I^*_5$, the system shows chaotic intermittency, \cite{CD}, generating a finite number of action potentials before returning to rest.

\subsection{Action potential solitary waves}

Since in the HH model, the action potentials propagate without attenuation, we can define the characteristic curve of the action potential by the condition $dV=0$. As $dV=\frac{\partial V}{\partial x}dx+\frac{\partial V}{\partial t}dt=0$, then $\frac{dx}{dt}=-(\frac{\partial V}{\partial t})/(\frac{\partial V}{\partial x})$, implying that action potentials may have a well defined speed. We can follow the action potential space-time evolution by taking the maxima of the functions $-V(x,t)$ as a reference point. We shall call the curves defined by the condition $dV=0$ the characteristic curves of the solution of the HH equations (\ref{eq10})-(\ref{eq11}).

\begin{figure}[h]
\begin{center}
\includegraphics[width= 0.8\hsize]{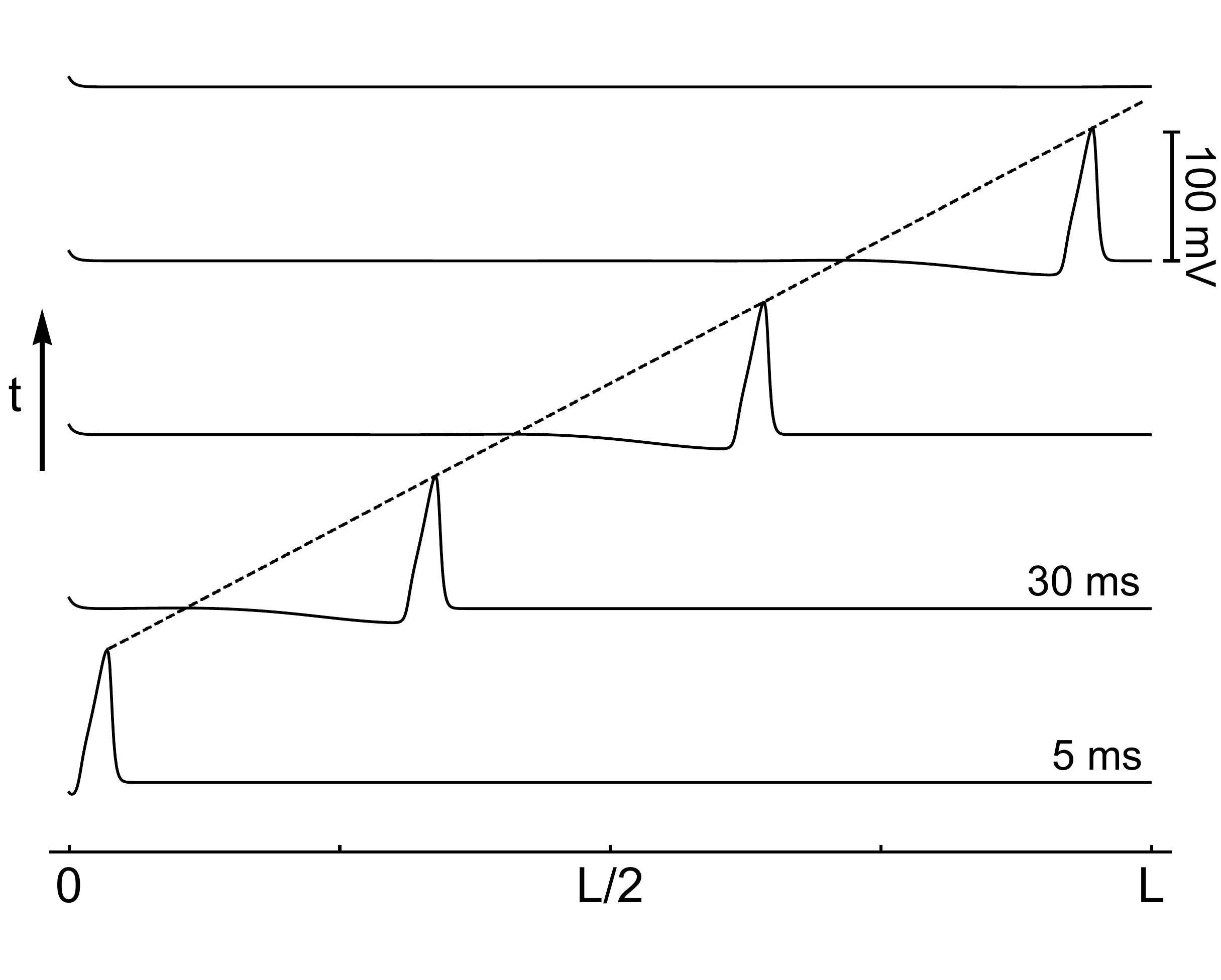}
\end{center}
 \caption{We show five time snapshots of the solutions $-V(x,t)$ of reaction-diffusion equations \eqref{eq10}-\eqref{eq11}. The bottom snapshot was taken at $t=5~$ms, and the succeeding snapshots differ by time intervals $\Delta t=25$~ms. We have considered that the axon is initially at the zero state $V(x>0,t<0)=0$ and the injected current at $x=0$ has the value $i_0=55.0$~$\mu$A/cm$^2$, during the entire simulation. The system is in the type I intermittency region (figure~\ref{fig:regions}), generating one spike. The dotted line shows a characteristic curve of the solutions of the HH equations. The propagation speed of the action potential spike is $v=12.14$~m/s. The action potential spike annihilates at the boundary $x=L$ without reflection.} 
 \label{fig:1spike}
\end{figure}

In figure~\ref{fig:1spike}, we show the solution of the HH system of equations (\ref{eq10})-(\ref{eq11}), responding to an injected current in the type I intermittency parametric region seen in figure~\ref{fig:regions}. In this case, a single action potential spike is generated at the injection point, propagating along the axon and disappearing at the boundary $x=L$. The characteristic curve of the solutions of the HH equations shows a linear profile, and the isolated action potential spike propagates along the axon at a constant speed. The slope of the characteristic curve translates to the propagation speed $v=12.14$~m/s.

\begin{figure}[h]
\begin{center}
\includegraphics[width= 0.8\hsize]{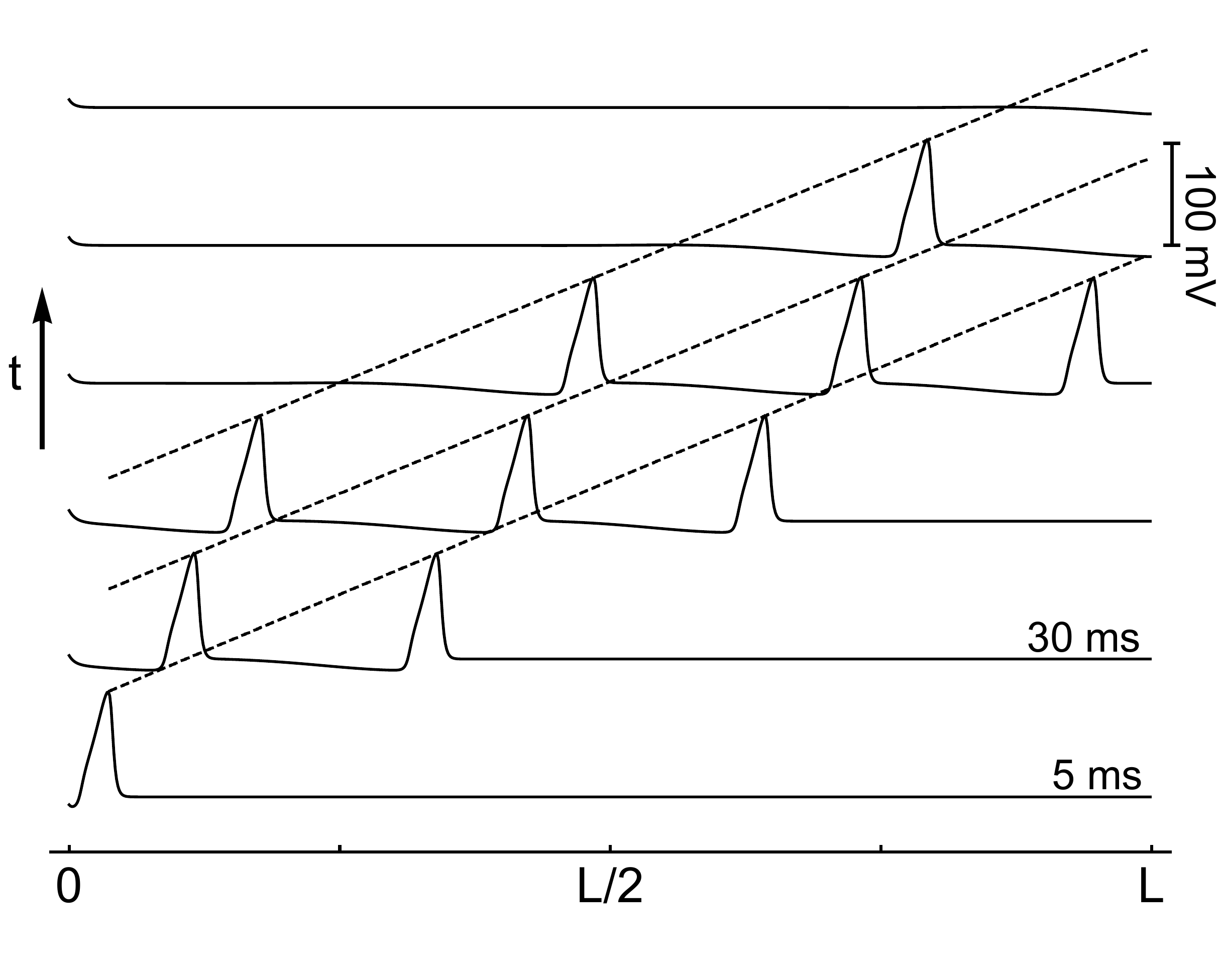}
\end{center}
\caption{Snapshots of the solutions $-V(x,t)$ of \eqref{eq10}-\eqref{eq11}. The conditions are the same as in figure~\ref{fig:1spike}, except that the injected current at $x=0$ is $i_0=57.7$~$\mu$A/cm$^2$ during the entire simulation. The system is in the type I intermittency region, generating three spikes. The dotted lines are three characteristic curves of the solutions of the HH equations. The speed of the front action potential spike is the same as in figure~\ref{fig:1spike}. During this observation time, the characteristic curves appear to be approximately parallel. Action potential signals annihilate at the boundary $x=L$.} 
 \label{fig:3spikes}
\end{figure}

In figure~\ref{fig:3spikes}, we show the solution of the HH system of equations (\ref{eq10})-(\ref{eq11}), in the region of type I intermittency, generating a total of three action potential sequential signals --- packet of spikes. This figure has been obtained for a larger value of $i_0$ compared with the simulations in figure~\ref{fig:1spike}. The speed of the first action potential spike is the same as in figure \ref{fig:1spike}. 

In both figures~\ref{fig:1spike} and~\ref{fig:3spikes}, the action potentials propagate without attenuation and annihilate at the boundary of the axonal domain. This effect is not observed with elastic waves. These action potentials propagate as solitary reaction-diffusion waves. After the annihilation of the action potentials at the boundary, the axon stays in a non-uniform and non-excitable state ($i_0\not= 0$). For a signal to propagate again along the axon, the electric state of the axon must return to the rest state $V=0$ and $i=0$, and the neurone may again be excited with a current above the threshold. 

\begin{figure}[h]
\begin{center}
\includegraphics[width= 0.8\hsize]{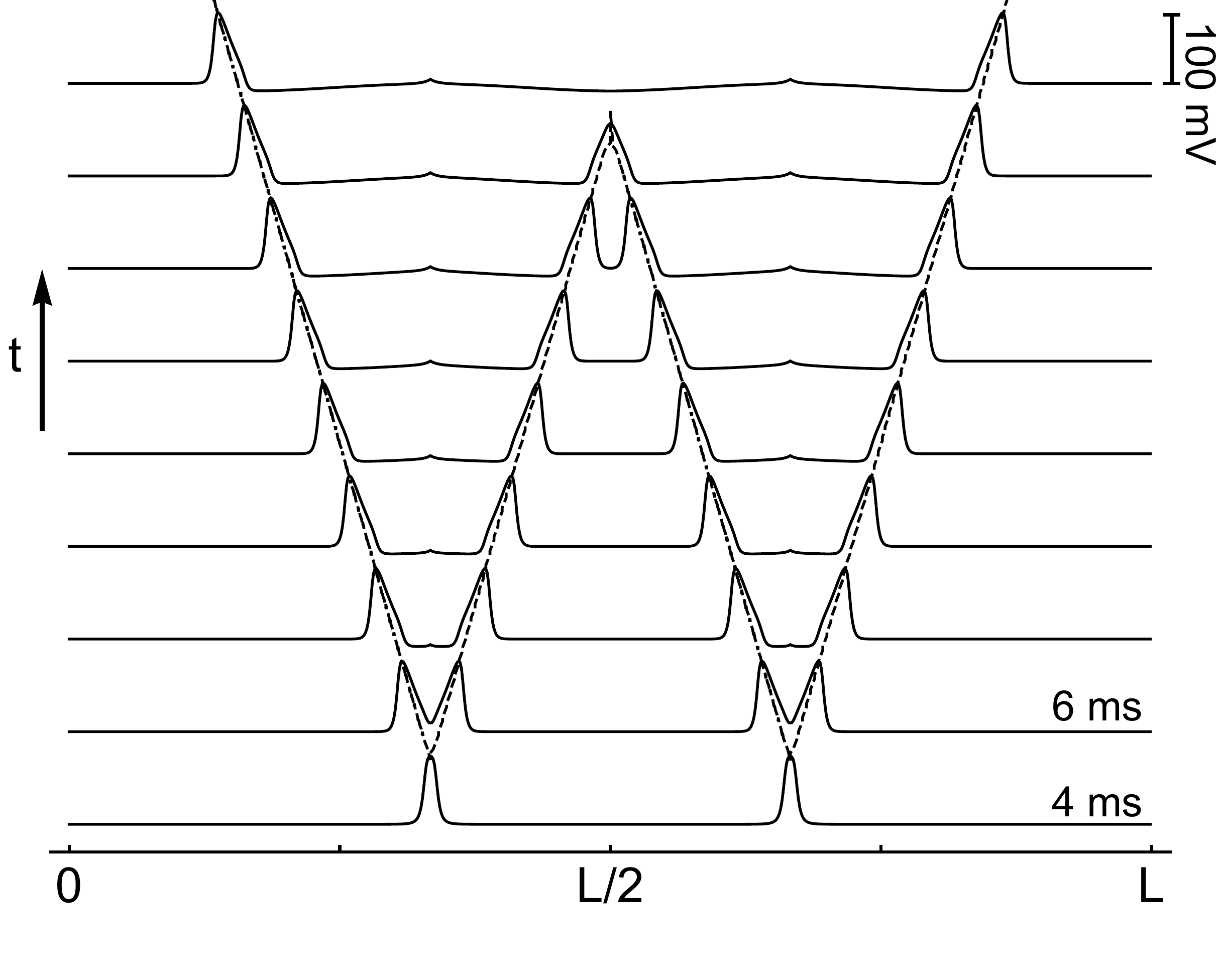}
\end{center}
 \caption{Snapshots of the solutions $-V(x,t)$ of \eqref{eq10}-\eqref{eq11}, for two injected currents at axon positions $x=L/3$ and $x=2L/3$. During the entire simulation, the injected currents have a value of $i_0=57.0$~$\mu$A/cm$^2$. The dotted lines represent the characteristic curves.} 
 \label{fig:spike_collision}
\end{figure}

In figure~\ref{fig:spike_collision}, we analyse the case where two current sources are injected into the interior of the axon at the longitudinal coordinates $x=L/3$ and $x=2L/3$. At each injection point, one action potential spike is generated, and, during propagation, each of them splits into two action potential spikes, moving in opposite directions. These four action potentials have the same amplitude as the action potentials in figures \ref{fig:1spike} and \ref{fig:3spikes}. Then, later, the two action potentials that travel towards the centre of the axon collide at $x=L/2$ and annihilate each other --- another effect characteristic of reaction-diffusion waves \cite{DV}, and \cite{SD}. The chosen injected current value is also within the type I intermittency region. After the collision at the boundaries, all the spikes disappear. The axon stays in a non-uniform and non-excitable state.

These numerical results show that the solutions of the HH equations do not propagate as elastic waves, as conjectured by Hodgkin and Huxley, but behave instead as reaction-diffusion waves. We have shown that the action potentials are not reflected at the axon boundary. Furthermore, the dynamic behaviour observed in figure~\ref{fig:spike_collision} is not compatible with that of elastic waves, where the amplitudes would be halved when the action potential splits in two, and the collision of two waves fronts would not result in annihilation.

\subsection{Action potential waves}

\begin{figure*}[!ht]
\centering
\includegraphics[width=\textwidth]{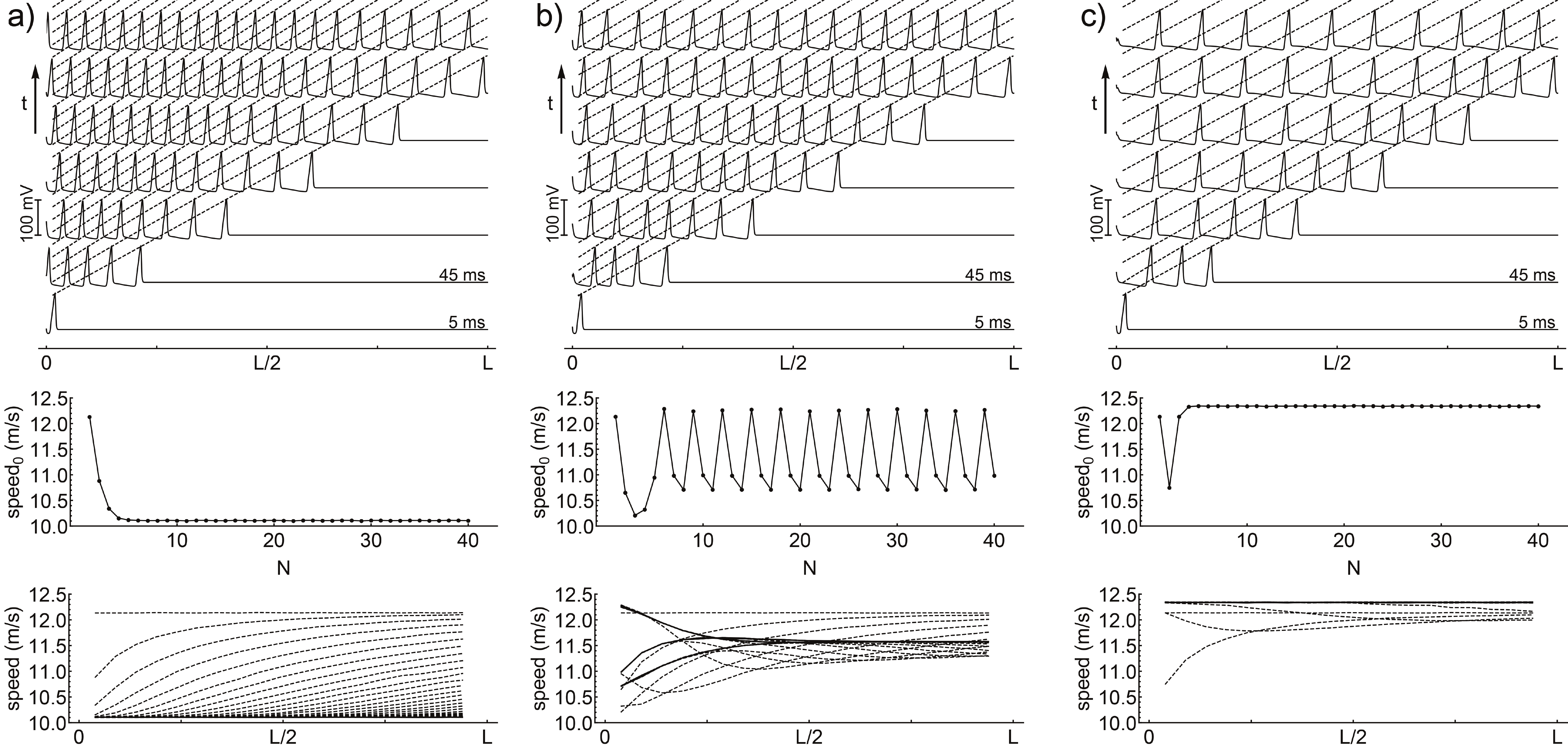}\\
 \caption{In the first row, we show snapshots of the solutions $-V(x,t)$ of \eqref{eq10}-\eqref{eq11}, for injected currents a) $i_0=200.0~\mu$A/cm$^2$, b) $i_0=300.0~\mu$A/cm$^2$, and c) $i_0=351.4~\mu$A/cm$^2$, at $x=0$ during the entire simulation. We also show some of the characteristic curves of the solutions. In the second row, we show the initial speed (for $x<<L$) of the first 40 action potentials generated by the system. The horizontal axis is the spike number. In the third row, we show the evolution of the speed of the first 40 action potentials as they propagate throughout the axon -- dashed lines correspond to spikes 1 to 29, and full lines to spikes 30 to 40. In a) and c), we have an asymptotically oscillatory response and, in b), a chaotic response.}
 \label{fig:oscillations}
\end{figure*}

In this section, we have extended the spatial domain of the simulation to L = 250 cm, discretised in M = 2000 small intervals.
In figure~\ref{fig:oscillations}, the axon is excited at $x=0$ with a large current $i_0$, in three different oscillatory regions of figure~\ref{fig:regions}. 

In figures \ref{fig:oscillations}a, the neurone is excited with current $i_0=200~\mu$A/cm$^2$, in the periodic oscillatory region (figure~\ref{fig:regions}). The initial speed of the action potentials converges quickly to a fixed value, as seen through $speed_0$. While this initial velocity converges, the action potential spikes accelerate while propagating through the axon. 

In figures \ref{fig:oscillations}b, the neurone is excited with current $i_0=300~\mu$A/cm$^2$, in the chaotic oscillatory region $[I^*_2,I^*_3]$. The initial speed of the action potentials converges to a period-3 solution. However, as the spikes propagate along the axon, this period-3 disappears, giving a constant propagation speed. The chaotic effect of this region in the propagation speeds of the action potentials is merely transient, dissipating as the spikes advance through the axon.

In figures \ref{fig:oscillations}c, the chosen current is $i_0=351.4~\mu$A/cm$^2$, in the small chaotic oscillatory region $[I^*_4,I^*_5]$. Even though this small region shows period bifurcations, which give way to a chaotic intermittency regime (as shown in \cite{CD}), the speed profiles do not give any hint of this. The simulations show that the initial velocity quickly converges to a fixed value as the action potentials travel along the axon. The chaoticity of signals is present in the non-uniformity of the distances between consecutive action potential spikes.

\begin{figure}[h]
\begin{center}
\includegraphics[width= 0.8\hsize]{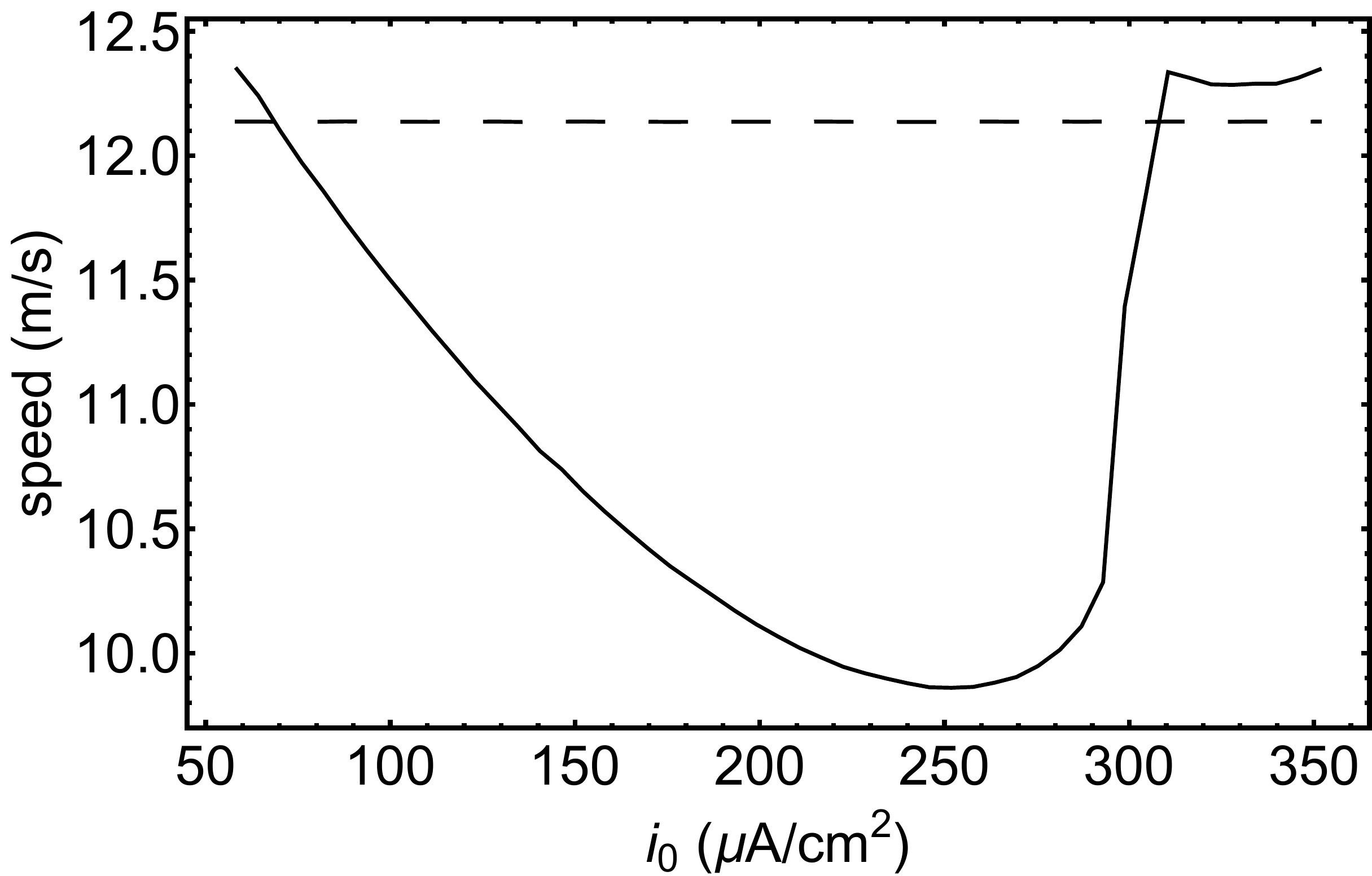}
\end{center}
 \caption{Action potential propagation speeds for the entire oscillatory region $[I^*_1,I^*_5]$. The dashed line represents the speed of the first action potential, which is always constant. The full line represents the final speed of the later action potential spikes (N$~> 30$), measured at the end of the axon, as in figures \ref{fig:oscillations}c).}
 \label{fig:velocities}
\end{figure}

In figure \ref{fig:velocities}, for different values of current $i_0$, we show the velocity profiles for the whole oscillatory region $[I^*_1,I^*_5]$.

The speed of the first action potential spike is the same ($v=12.14~$m/s), regardless of the value of injected current. We had already observed the same value for the type I intermittency region in figures \ref{fig:1spike} and \ref{fig:3spikes}. On the other hand, the final speed varies as the injected current increases. At first, the velocity decreases, reaching its lowest value around 250 $\mu$A/cm$^2$. Then, it increases until $I^*_2$, where transient chaos influences the period and speed with which the spikes are generated (as shown in figure \ref{fig:oscillations}b). In $[I^*_3,I^*_4]$, the final velocity of the system stabilises, only varying slightly as the current keeps increasing until it stops existing at the end of the oscillatory region.

\begin{figure}[h]
\begin{center}
\includegraphics[width= 0.8\hsize]{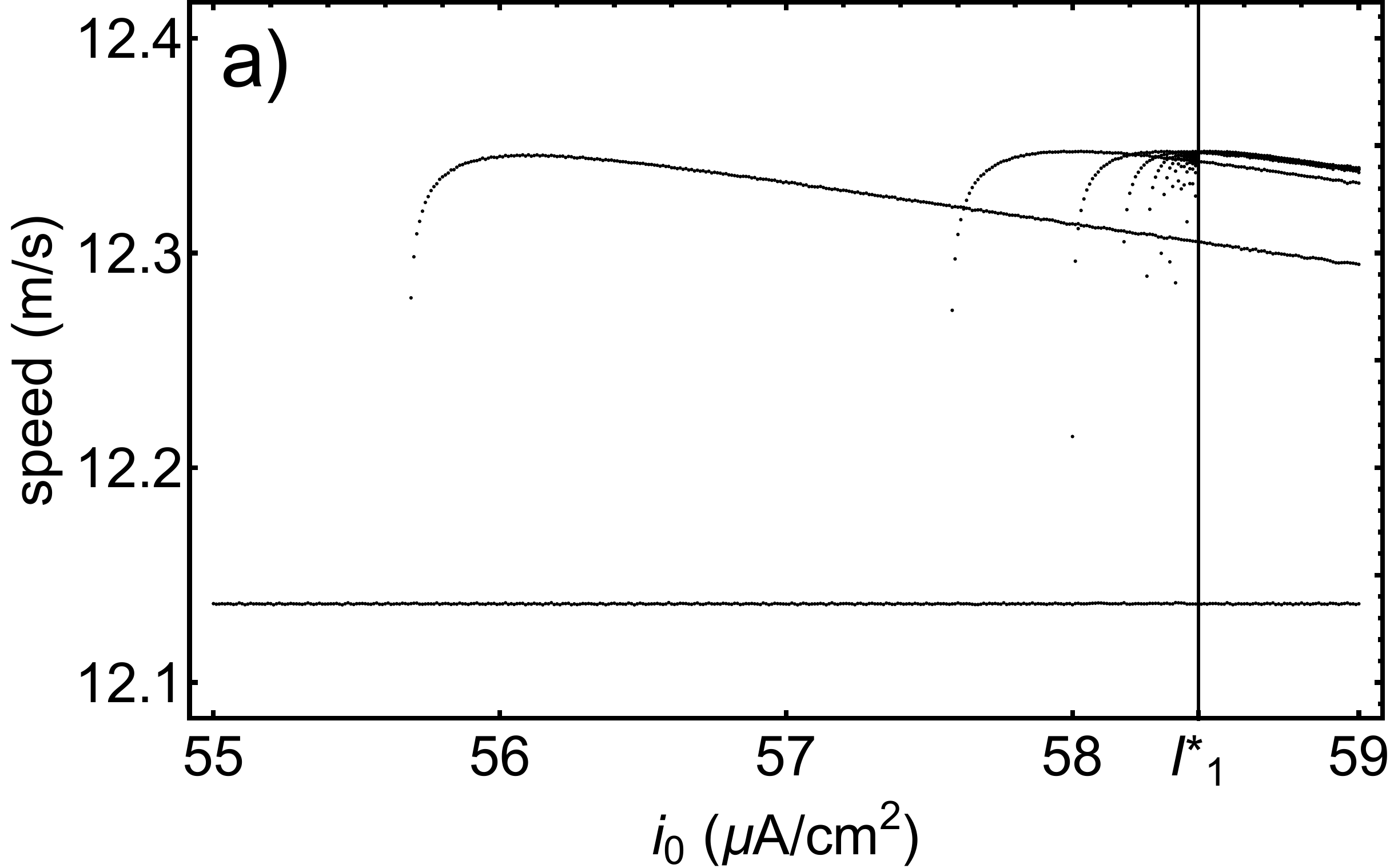}
\includegraphics[width= 0.8\hsize]{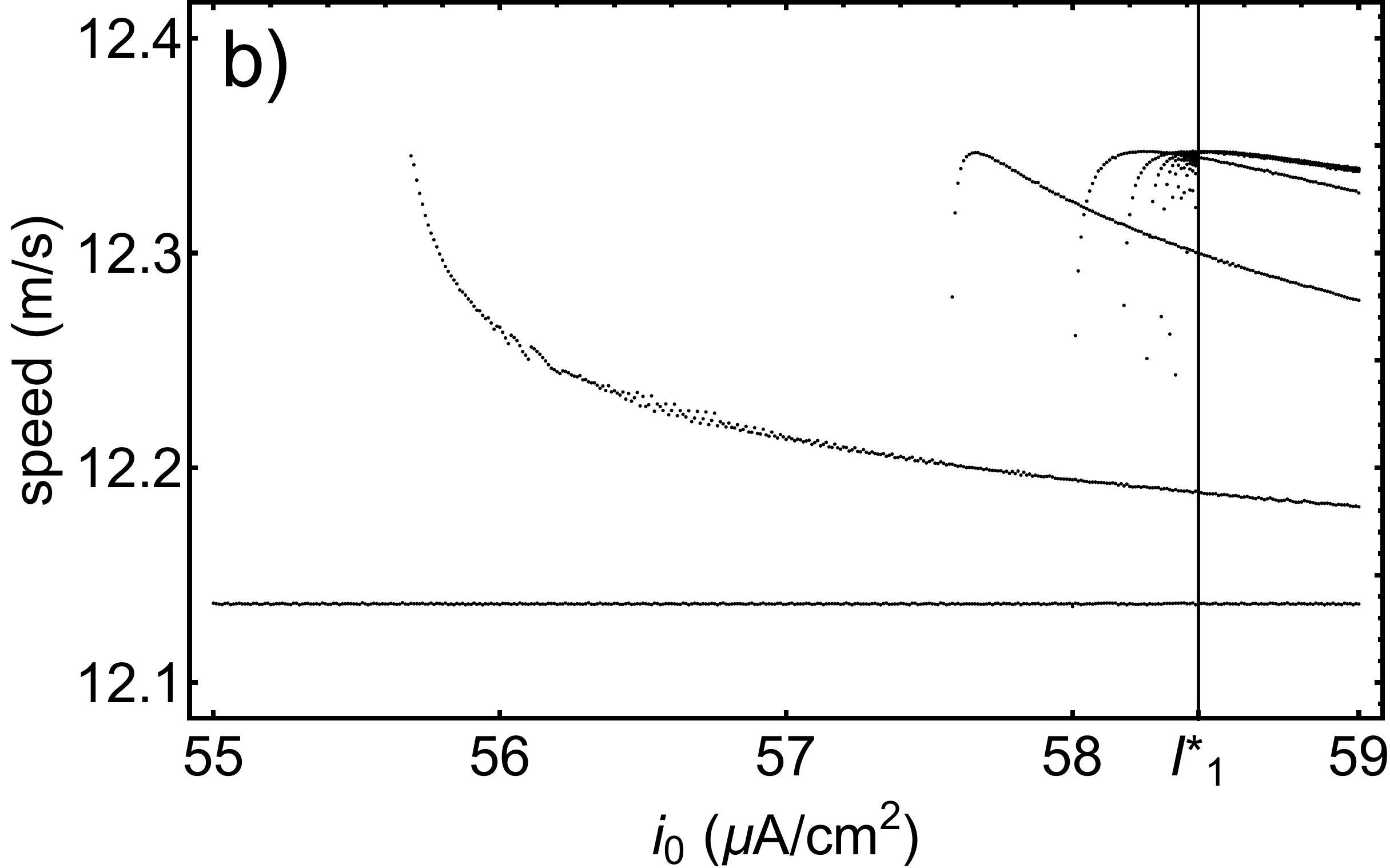}
\end{center}
 \caption{Propagation speeds of action potential spikes in the vicinity of $I^*_1$. The horizontal line corresponds to the velocity of the first spike, which remains constant throughout the axon (see figure \ref{fig:oscillations}). In a), the speed of the action potentials is measured at the beginning of the axon; in b), it is measured at the end of the axon. All of the dots correspond to the measured velocities of the action potentials. The transition from type I intermittency to periodic oscillations occurs at  $i_0=I^*_1$. 
 }
 \label{fig:v_typeI}
\end{figure}

In figure \ref{fig:v_typeI}, we  measured the velocities of the action potentials in the type I intermittency region, and we analysed how their profile changes as the transition to the oscillatory region occurs.

 For low values of current ($i_0 \le 55.5$~$\mu$A/cm$^2$), only one action potential spike is generated. For  $i_0 = 55.0$~$\mu$A/cm$^2$, the speed is  $v=12.14$~m/s. 
 The speed of the first action potential spike remains constant during propagation, both in the intermittency and oscillatory regions. This is consistent with what we have seen in figures \ref{fig:1spike}, \ref{fig:3spikes}, \ref{fig:oscillations}, and \ref{fig:velocities}. 
 As the current increases ($ 55.5 < i_0 < I^*_1$), we can see additional action potential spikes with different speeds. At $i_0 = 57.7$~$\mu$A/cm$^2$, we can distinguish three different propagation speeds, corresponding to the three action potentials spikes seen in figure \ref{fig:3spikes}, where the same current was injected. Whereas in figure \ref{fig:3spikes} all of the characteristic curves seemed to be linear, the speed of the second and third spikes do not remain constant during the propagation, having different values at the beginning (figure \ref{fig:v_typeI}a) and end (figure \ref{fig:v_typeI}b) of the axon. As the current approaches $I^*_1$,  the number of spikes increases  (an effect of type I intermittency \cite{CD}), and the velocity profile approaches the profile of the oscillatory region.

\begin{figure}[h]
\begin{center}
\quad \includegraphics[width= 0.76\hsize]{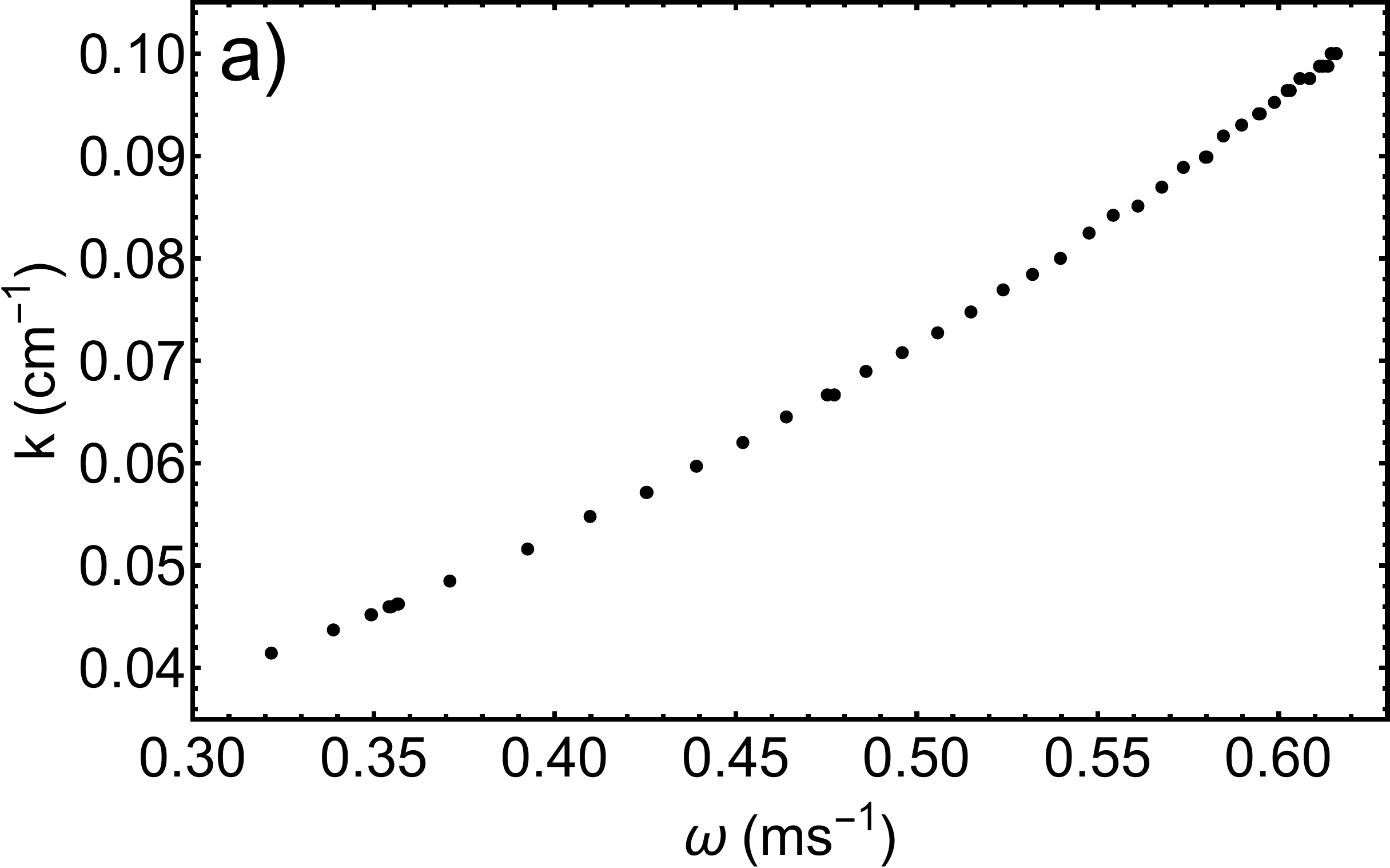}
\includegraphics[width= 0.8\hsize]{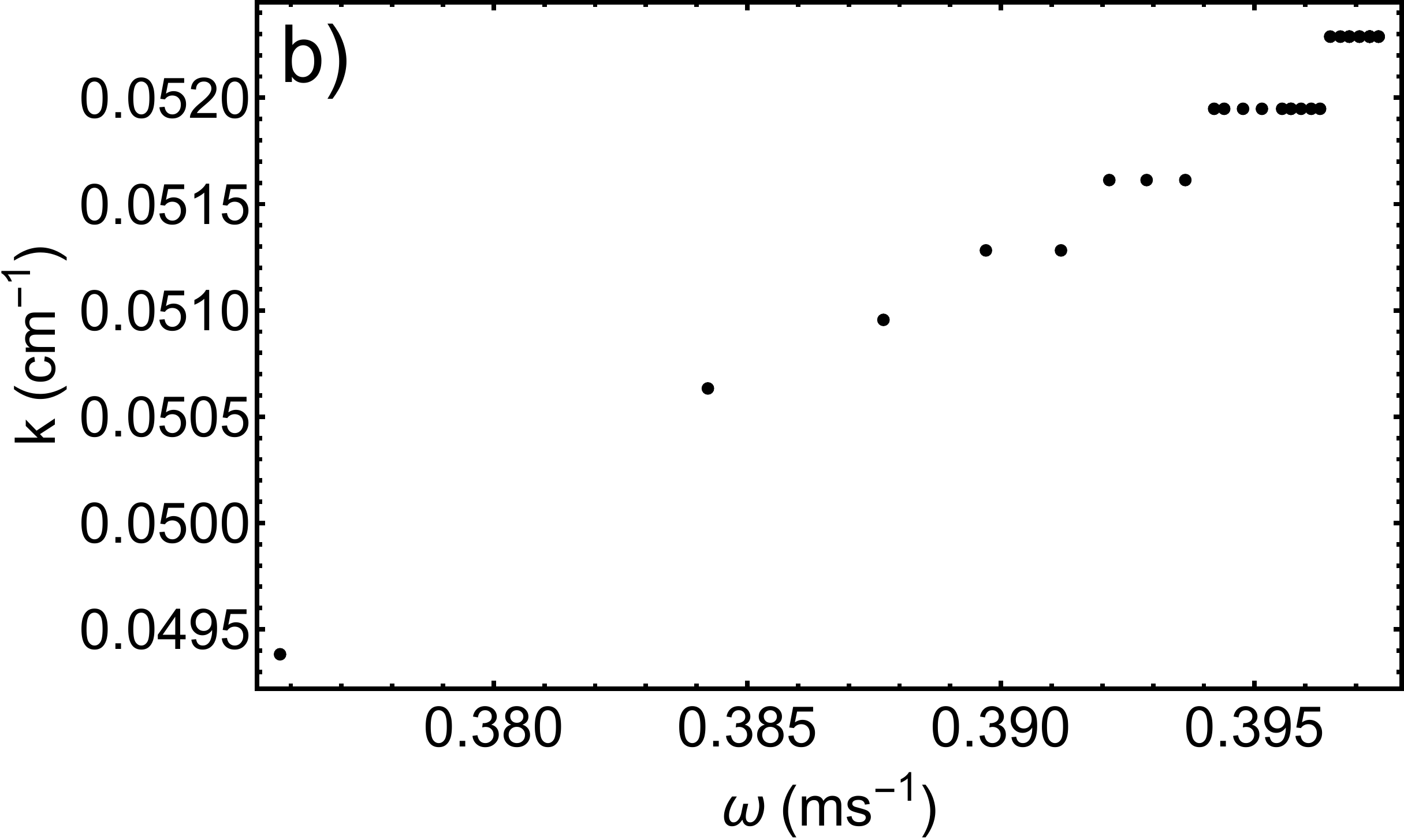}
\end{center}
 \caption{Asymptotic dispersion relations for the oscillatory region $[I^*_1,I^*_4]$. The measurements were made in the last quarter of the axon. In a), the period ($T=1/\omega$) and wavelength ($\lambda=2\pi/k$) were calculated through spikes 30 and 31. In b), spikes 1 and 2 were used.} 
 \label{fig:disp_relation}
\end{figure}

In figure \ref{fig:disp_relation}, we have calculated the asymptotic dispersion relation for the oscillatory region $[I^*_1,I^*_4]$. In a), we have calculated the period and wavelength between spikes 30 and 31 at the last quarter of the axon. In b), we have calculated the dispersion relation at the same location in the axon, but between spikes 1 and 2, before the system has reached the asymptotic regime. This shows that the asymptotic dispersion relation of the oscillatory regime of the HH equation \ref{eq1} depends on the position of the spikes along the axon.

\section{Final remarks and conclusions}\label{sec4}

To test the predictions of the HH model (\ref{eq1}) with patch-clamp data and without the Hodgkin and Huxley assumptions (\ref{eq2}) and (\ref{eq3}),  the first requirement is to measure the current as a function of the spatial position along the axon. Due to the diffusive nature of the current propagation along the axon, the second requirement is to test if action potential spikes propagate without attenuation. Without fulfilling these requirements, the HH model can not be validated.

Another important prediction of the HH model (\ref{eq1}) about the propagation of axonal signals is the existence of the type I intermittency phenomenon associated with a saddle-node bifurcation of limit cycles. This bifurcation is tuned by the magnitude of the applied current $i_0$ to the axon. 
Let $I^*_1$ be the $i_0$ bifurcation value. 

For $i_0<I^*_1$, and depending on the electrophysiological state of the axon (transmembrane potential, ionic concentrations, etc.), we may have no spikes at all, one spike or several spike responses up to some maximum number $N$. This number $N$ relates with $I^*_1$ and $i_0$ through the relation $\ln N=C-(\ln (I^*_1-i_0))/2$, characteristic of type I intermittency, \cite{CD}. This multi-spike phenomenon has never been reported in voltage-clamp experiments,  \cite{Cl}. On the other hand, its observation would corroborate the existence of the bifurcation predicted by the HH model. If this behaviour is not observed, the HH model validity would be for values of $i_0$   below $I^*_1$. In this case, the HH model equations have a unique stable steady state associated with the equilibrium values of the Nernst potentials, and the action potential signal could not exist. In fact, action potentials are due exclusively to the type I intermittency phenomenon.

The importance of the existence of a saddle-node bifurcation of limit cycles implies that axonal signals may respond to external stimuli with an approximately periodic or even chaotic response. The observation of this type of response is an essential biological phenomenon predicted by the HH model.

Even in the case of negative observations of the several spiky phenomena and associated intermittency, it would be essential to observe the possibility of propagation of signals in the two opposite directions of the axon. This phenomenon is believed to occur, \cite{Oh}, as well as the annihilation when two isolated action potential spikes collide. These are intrinsic phenomena associated with the nature of reaction-diffusion equations and the HH model. If these interaction patterns fail, we can not say that action potentials are reaction-diffusion waves. In this case, the derivation of a more detailed model for studying electric phenomena in cells and axons should be reconsidered.

Extensive simulations have shown that time intervals between consecutive action potential spikes vary during the time and along with the axon position, which is characteristic of the reaction-diffusion nature of the HH model.
{
\section*{Acknowledgments}
RD would like to thank IH\'ES for hospitality and Simons Foundation for support. 
}

\end{document}